\documentclass[12pt]{article}

\usepackage{ucs}
\usepackage[utf8x]{inputenc} 

\usepackage{graphicx}
\usepackage[authoryear]{natbib}

\usepackage[margin=1in]{geometry}
\usepackage{amsmath}
\usepackage{amssymb}

\newcommand{\Ei}{\text{Ei}}
\newcommand{\Chi}{\text{Chi}}

\usepackage{setspace}

\begin{document}

\title{The look-ahead effect of phenotypic mutations}

\author{~\\
\\
\\
\\
 \\
Dion J. Whitehead$^{*1}$, Claus O. Wilke$^{*2}$, David Vernazobres$^{1}$, Erich Bornberg-Bauer$^{1}$}

\date{~\today}

\maketitle
~\\
\\
\\
\\
\\
\\
\\
\\
\\
\\
$^*$ Both authors contributed equally to this work.
\\
\\
$^1$Institute for Evolution and Biodiversity,
The Westphalian Wilhelms University of Muenster,
48149 Muenster, Germany

\noindent $^{2}$Center for Computational Biology and Bioinformatics and Institute for Cell and
Molecular Biology, Section of Integrative Biology, University of Texas,
Austin, Texas 78712

\maketitle

\clearpage

Running Head: Look-ahead effect phenotypic mutations \\


Corresponding author:

Dion J. Whitehead

Institute for Evolution and Biodiversity, 

The Westphalian Wilhelms University of Muenster, 

Schlossplatz 4, D48149 

Germany

Email: dion@uni-muenster.de

Phone: +49-(0)251-83-21633

Fax: +49-(0)251-83-21631



\clearpage
\doublespacing
\begin{abstract}
The evolution of complex molecular traits such as disulphide bridges often requires multiple mutations.  
The intermediate steps in such evolutionary trajectories are likely to be selectively neutral or deleterious.  Therefore, large populations and long times may be required to evolve such traits. 
We propose that errors in transcription and translation may allow selection for the intermediate mutations if the final trait provides a large enough selective advantage.  We test this hypothesis using a population based model of protein evolution.
If an individual acquires one of two mutations needed for a novel trait, the second mutation can be introduced into the phenotype due to transcription and translation errors.
If the novel trait is advantageous enough, the allele with only one mutation will spread through the population, even though the gene sequence does not yet code for the complete trait.  
The first mutation then has a higher frequency than expected without phenotypic mutations giving the second mutation a higher probability of fixation.  Thus, errors allow protein sequences to ''look-ahead'' for a more direct path to a complex trait.  
\end{abstract}

\clearpage

\section{Introduction} 
According to a central principle of molecular evolution, the likelihood that a given mutation occurs is independent of the mutation's phenotypic consequences. Organisms cannot choose specific  mutations.  This tenet was challenged by \cite{Cairns-1988}, who observed that under a certain selective pressure, \emph{E. coli} cells appeared to acquire an excess of beneficial mutations.  The idea that cells can somehow `direct' evolution was thought provoking, and stimulated many investigations (for reviews see \cite{Bridges-1998, Foster-1999, Cairns-1998, Hall-1998, Rosenberg-2001}). 
While the notion that cells can directly decide in which genomic regions to increase their mutation rate has been mostly abandoned \citep{Foster-1998, Cairns-1998}, the original observations by \cite{Cairns-1988} have been corroborated (see above reviews).

If mutations arise independently of their phenotypic consequences, then how can 
adaptations occur that require multiple amino acid mutations and for which the intermediate stages are either selectively neutral or disadvantageous?  Large populations can climb multiple fitness peaks, even with disadvantageous intermediate alleles \citep{Behe-2004, Weinreich-2005}.  
Although no new mechanisms are therefore required to explain the evolution of complex proteins \citep{Lynch-2005}, we propose that errors in transcription and translation (\emph{phenotypic} mutations) allow the selection of the intermediate mutations of a multiple-mutation requiring trait, and can thus speed up the evolution of complex traits.


Studies on the phenotypic mutation rate indicate that it is orders of magnitude larger than the genotypic mutation rate \citep{Springgate-1975, Edelmann-1977}: the global phenotypic non-synonymous mutation rate has been estimated to be $4.5 \times 10^{-4}$ mistranslations per codon \citep{Ellis-1982,Shaw-2002}, compared with a genotypic mutation rate of between $\sim 10^{-7}$ to $10^{-11}$ \citep{Drake-1998}.  Consequently, for a protein of 300 residues, on average more than 1 in 10 copies of the protein will contain a mutation.  
Using mutation rates derived from the literature and conservative biological assumptions, we show via mathematical modeling and simulations that phenotypic mutations allow evolution to select for neutral intermediate alleles of a multi-mutation trait, actually selecting for proteins whose exact DNA sequence is not in the organism under selection.  Evolution is then able to look ahead for evolutionary jackpots in sequence space.

Our theory is based on the following hypothetical scenario. A protein can increase the fitness of an individual if it evolved a specific trait.  This trait requires two mutations, for example a disulphide bridge between two cysteine residues. Having only one of the required mutations is either selectively neutral or deleterious, however when an individual has only one mutation, small amounts of the protein with both mutations will be produced due to phenotypic mutations.  
If the presence of both mutations at low concentrations provides even a small fitness improvement then 
the allele with one mutation will spread though the population. 
As the frequency of the intermediate allele increases, there is a greater probability that if the second mutation occurs, it will be the presence of the first mutation, and thus provide the full fitness benefit.

The aim of this article is to show that adaptive phenotypic mutations can undergo positive selection under biologically plausible conditions, allowing proteins greater access to features involving multiple mutations.  
We thus introduce the notion of  a ''look-ahead'' effect. The name indicates that seemingly unsurmountable evolutionary barriers can be overcome thanks to phenotypic mutations which are not yet present in the genome.  We wish to emphasize that the look-ahead concept is firmly grounded on the idea of chance and necessity and by no means insinuates a teleological feature
of molecular evolution.

\section{Materials and Methods}


\begin{table}[th]
\begin{tabular}{ll}
\multicolumn{2}{c}{\textbf{Variables used in this work}} \\
&\\
\hline
$\beta$ & Phenotypic mutation rate from allele 1 to allele 2\\
$\mu$ & Null mutation rate per residue\\
$N$ & Population size\\
$r$ & Number of residues available for the second mutation\\
$s$ & Selection coefficient for allele 2\\
$U$ & Generational (DNA) mutation rate from allele 1 to allele 2\\
\hline
\end{tabular}
\end{table} 

\paragraph{Model assumptions:}
We model the scenario of a protein evolving a trait that requires two mutations.  The model is based on a population-genetics framework where a single gene can evolve into different alleles.  We do not consider duplication and divergence of genes.  In addition, the process described here will likely only occur for proteins with sufficiently long half-lives, as the protein must persist for some time to exert a phenotypic effect.  As we model only a single gene, we expect our results to be more relevant for single-celled organisms and viruses than for multicellular organisms, which tend to have larger genomes and smaller effective population sizes than microorganisms.

\begin{figure}[htbp]
\centering
\includegraphics[width=120mm]{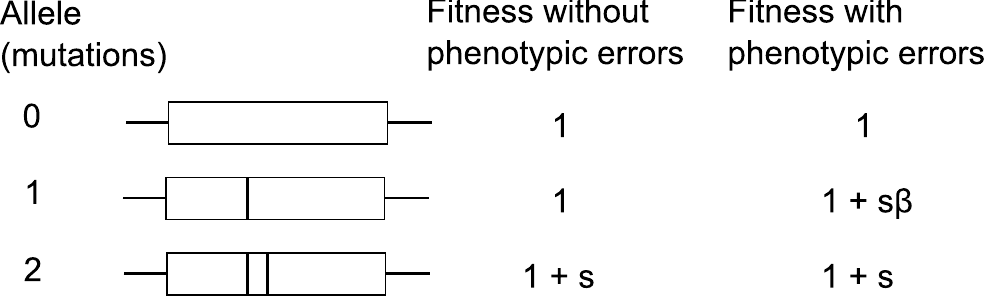}
\caption{The three alleles (or genotypes).  The vertical lines in the genes indicate the number of key mutations required for the novel two-residue function.  The fitness of the allele 1 increases if phenotypic mutations are taken into consideration.}
\label{fig:diagram}
\end{figure}

The model consists of the evolution of three non-recombining haploid genotypes, where each genotype contains one of the three alleles shown in Figure \ref{fig:diagram}.  
The three different alleles are named according to number of relevant mutations, corresponding to zero mutations (allele 0), a single mutation (allele 1), and both mutations (allele 2) required for the adaptive feature.  Having both mutations of the adaptive feature provides a selective advantage $s$.  We assume that the intermediate allele (allele 1) is selectively neutral \emph{if transcribed and translated without error}.  We specifically take into consideration errors in transcription and translation, that is, \emph{phenotypic mutations}.

In the model, the population initially consists of one individual carrying allele 1 and $N-1$ individuals carrying allele 0.  
So long as allele 1 is present, allele 2 can be generated by mutations.  The population evolves for a fixed time period, during which allele 2 can be generated by mutation and go to fixation.  

In each generation, selection increases the frequency of the alleles according to their corresponding fitness values.  
Allele 0 has a fitness of 1.  Allele 2 has a fitness of $1+s$, where $s$ is the selection coefficient provided by the adaptive feature.  The fitness of allele 1, the intermediate allele with only a single mutation, depends on the phenotypic error rate.  
Most phenotypic errors will be neutral or deleterious, however some will be beneficial. 
For simplicity, we assume that the length of the protein and the expression level are both constant.  In addition, we also assume that the cost due to deleterious phenotypic errors is also constant.  The effect of these parameters on this model will be the subject of future work.  
If there are no phenotypic mutations, allele 1 has the same fitness as allele 0.  However, if phenotypic mutations occur, allele 1 can produce a small number of allele 2 proteins due to phenotypic errors. The fitness of allele 1 is therefore dependent on the number of such errors.

We assume that fitness is a linear sum of individual proteins, meaning that if some phenotypic variants of a protein have a higher fitness, then the overall fitness of that allele is proportionally increased.  

We let $r$ be the number of residues that can potentially complement the first mutation to provide the full two-residue adaptive feature.  These $r$ residues represent, e.g.,\ the sites at which the second cysteine of a cysteine bridge could arise; other similar two-residue mutations that significantly improve functionality can be proposed.  
Two residues that comprise an adaptive trait are likely to co-evolve, because if a mutation occurs in one of the residues, selection strongly favors a compensatory mutation in the other.  
Based on a large data set, \citet{Martin-2005} found that co-evolving residues are spatially near.  
Co-evolving residues were, on average, 98.6 amino-acids apart along the sequence, but had a mean spatial distance
of 6.9 \AA.  
This spatial distance can be compared to the width of the van der Waals volume of an amino-acid (5-6 \AA), showing that most co-evolving residues are effectively in contact proximity.  
Therefore, $r$ is mostly independent of the size of the protein, as long as the protein is of sufficient length.
\citet{Bloom-2006} calculated the mean contact density (the mean number of residues in contact with a given residue) for 194 yeast proteins, and found that most residues have a mean contact density of seven to eight residues. In this work we use $r=8$.

Given $r$ possible positions for the second residue, and assuming that each position requires a specific residue, the fraction of proteins of allele 1 containing the second (now highly beneficial) mutation is $\beta=\frac{r}{19} \lambda$, where $\lambda$ is the per codon non-synonymous phenotypic mutation rate.  In this model, we use $\lambda=4.5 \times 10^{-4}$ mistranslations per codon \citep{Ellis-1982,Shaw-2002}.  The fraction $\beta$ of allele-2 proteins  contribute to the fitness, giving allele 1 a fitness of $1+s \beta$.

When considering genetic (i.e. inherited) mutations, for simplicity we neglect back mutations (e.g.\ from allele 1 to allele 0), and assume there are no recurrent mutations of allele 1 from allele 0 (the model starts with a single copy of allele 1).  Allele 2 arises via a mutation from allele 1. We ignore the possibility of a double mutation directly from allele 0 to allele 2, as this probability is extremely small.
The genetic mutation rate for allele 1 mutating into allele 2 is derived as follows:  For microbes, the rate of mutations per nucleotide per generation is between $\sim 10^{-7}$ to $10^{-11}$ \citep{Drake-1998}.  Here we use $10^{-8}$ as the non-synonymous mutation rate per codon per generation.  The resulting mutation rate for changing allele 1 into allele 2 is $U=\frac{r}{19} 10^{-8}=\frac{8}{19} 10^{-8}$.  


Genes can also acquire null mutations, rendering the gene non-functional and therefore eliminating the organism.  The null mutation rate for protein-encoding genes is on the order of $10^{-6}$ per generation \citep{Drake-1998}. However, this rate will depend on the length ($L$) of the protein.  Assuming an average protein length of 300 residues, the per-residue null mutation rate is given by $10^{-6}/300 = \sim 3.3^{-9}$.  For a protein of length $L$, the null mutation rate is given by $\mu=3.3^{-9} L$.

\paragraph{Simulations:} The numerical simulations were written in Java using the Colt scientific library \citep{Colt-2004} for the generation of random numbers.  The analytic expressions were evaluated using both Mathematica and Python, the latter in conjunction with the SciPy package \citep{SciPy-2007}.  Source code for the numerical simulations is available on request from DJW.

The population in each simulation is represented by three numbers, corresponding to the abundance  of each of the three alleles.  As described, the initial abundances are $N-1, 1, 0$ for alleles $0, 1, 2$, respectively. The simulation runs for a specified number of generations $T$. We used $T=5 \times 10^5$ throughout this work.  Strictly speaking, $T$ is the number of generations in which allele 1 can mutate into allele 2; for later generations this possibility of mutation is disabled.  If allele 2 is present at time $T$, then the simulation is continued until allele 2 is either lost or has reached fixation.
Generations are discrete, with mutations, selection, and drift occurring at each generation.  During each generation we perform the following steps.  First we check if either allele 0 or allele 2 has reached fixation; if so, we stop the simulation, as both cases are absorbing states.  Next, for each allele we check for null mutations by drawing a random number from the Poisson distribution where the expected number of events is the null mutation rate $\mu$ multiplied by the total number of individuals with the given allele.  Mutations from allele 1 to allele 2 are computed in a similar manner, where the expected number of events is $U$ multiplied by the number of allele 1 individuals.  Then, after the possible production of the mutant allele 2, selection acts on the fitness of the alleles, where the frequency of each allele is multiplied by its corresponding fitness, $[1, 1 + s\beta, 1+s]$ for alleles $[0,1,2]$, giving the new number of alleles in a possibly larger population.  Finally, the next population of $N$ individuals is chosen by recursively sampling from the binomial distribution, representing random genetic drift.  Allele 0 is first sampled with the mean=(frequency of allele 0), and the (number of trials)=$N$.  Allele 1 is then sampled from the combined allele 1 and 2 individuals.  The number of simulations where allele 2 becomes fixed is divided by the total number of simulations, giving an estimate of the fixation probability.  The number of simulations for each parameter set was between $10^8$ and $10^9$.

\section{Results}
\subsection{Analytical fixation rate of allele 2}

To calculate the fixation rate of allele 2 we have to consider the two fates of allele 1.  Firstly, allele 1 can become lost.  In this case allele 2 can only be generated during the period of drift of allele 1.  The alternative fate of allele 1 is fixation.  Then allele 2 can be generated either while allele 1 drifts to fixation or after allele 1 is already fixed.  We would like to know how many mutation events from allele 1 to allele 2 are expected for either fate of allele 1.  We let $n(s\beta)$ be the expected number of mutation events for when allele 1 is eventually fixed, and $n_{loss}(s\beta)$ be the expected number of mutation events for the case when allele 1 is lost.  We can calculate $n(s\beta)$ and $n_{loss}(s\beta)$ from diffusion theory, by integrating over the sojourn times of allele 1. The corresponding calculations are cumbersome but straightforward, and for the sake of brevity we present the details in the Appendix (\ref{number-of-mutations-conditional-on-extinction} and \ref{mutations-given-time-interval}).  For $n(s\beta)$, allele 2 can be generated as allele 1 drifts to fixation, and also after allele 1 has already reached fixation.  For $n_{loss}(s\beta)$, allele 2 can only be generated while allele 1 drifts.  

Assuming that $m$ is the expected number of times allele 2 is generated, what is the probability that at least one copy goes to fixation? The probability of fixation of a single copy of allele 2 is $u(s)$ \citet{Kimura-1962}. (In Appendix \ref{probability-of-fixation}, we reproduce the exact expression for $u(s)$, as well as approximations for large and small $s$.) Thus, if allele 2 is generated $k$ times, its probability of fixation is $1-[1-u(s)]^k$. Since the probability that allele 2 is generated $k$ times follows a Poisson distribution with mean $m$, we find for the probability $v$ that at least one of the mutations to allele 2 goes to fixation 
\begin{align}\label{eqn:v}
  v &= 1-\sum_k \frac{m^k}{k!}e^{-m} [1-u(s)]^k\notag\\
    &= 1-e^{-mu(s)}.
\end{align}

We calculate this probability separately for $n(s\beta)$ and $n_{loss}(s\beta)$, setting $m$ equal to either of these values.  We assume that $T$ is sufficiently large so that allele 1 has time to reach fixation within this interval (we assume $T\gtrsim 2N$).  Then the probability $u_2(s,\beta)$ that allele 2 is generated and goes to fixation (starting with a single copy of allele 1) is
\begin{equation}\label{u2generalT}
  u_2(s,\beta) = u(s\beta)\big(1-e^{-n(s\beta)u(s)}\big) + \big(1-u(s\beta)\big)\big(1-e^{-n_\text{loss}(s\beta)u(s)}\big).
\end{equation}
The first half of the equation stems from the case when allele 1 eventually reaches fixation, where the probability that allele 1 becomes fixed, $u(s\beta)$, is multiplied by the probability $v$ that at least one copy of allele 2 is generated and fixed.  
The second half corresponds to the case of loss of allele 1 from the population, where the probability of loss of allele 1, $(1-u(s\beta))$, is multiplied by the probability of at least one mutation from allele 1 to allele 2 and subsequent fixation of allele 2.   Taking into account allele 2 mutations during allele 1 loss is important especially for small $s$.  Allele 1 is more likely to be lost than fixed for small $s$, but can occasionally drift for long times before being lost.

In the limit $\beta\rightarrow 0$, i.e., in the absence of phenotypic mutations, we find with Eqs.~\eqref{ufixsmall}, \eqref{nlosssmall}, and \eqref{nsmall}
\begin{equation}\label{u2fullneutralT}
  u_2(s,0) = \frac{N+1}{N}- e^{-NU(T-N)u(s)}/N-e^{-NUu(s)}.
\end{equation}
(We assume that $N\gg1$, and neglect corrections of order 1 compared to $N$.  Note that we cannot simplify
$(N+1)/N$ to 1, because for small $U$, $1-e^{-NUu(s)}$ and $(1-e^{-NU(T-N)u(s)})/N$
are of the same order in $N$.)

As we are interested in the effect of phenotypic mutations ($\beta > 0$) compared to the case without phenotypic mutations ($\beta=0$), we define the increase in the probability of fixation from advantagous phenotypic mutations (the look-ahead effect) as
\begin{equation}\label{eqn:xi}
  \xi = \frac{u_2(s,\beta)}{u_2(s,0)}.
\end{equation}

We can broaden the assumption of $T\gtrsim 2N$ to $T\rightarrow\infty$ with good accuracy. For $T\rightarrow\infty$, if allele 1 is destined to reach fixation, then the probability of generating at least one copy of allele 2 that goes to fixation approaches 1. Therefore, $1-e^{-n(s\beta)u(s)}\rightarrow 1$, in this limit, and thus
\begin{equation}
 \xi \approx \frac{\displaystyle u(s\beta) + \big(1-u(s\beta)\big)\big(1-e^{-n_\text{loss}(s\beta)u(s)}\big)}{(N+1)/N -e^{-NUu(s)}}.
 \label{eqn:final_xi}
\end{equation}
Apart from a correction for the case when allele 2 occurs while allele 1 is destined for extinction, Equation \eqref{eqn:final_xi} is just the ratio of the probability of allele-1 fixation in the presence and absence of phenotypic mutations, $u(s\beta)/u(0)=Nu(s\beta)$.

To first order in $s\beta$, Eq.~\eqref{eqn:final_xi} simplifies to (Appendix \ref{xi-smalls})
\begin{equation}
  \xi \approx 1+N s\beta + {\cal O}(s^2\beta^2).
  \label{eqn:xi-approx1}
\end{equation}
We can see from this equation that the look-ahead effect becomes important when $N$ is on the order of $1/(s\beta)$.

For $Ns\beta\gg1$, only the first term contributes to the numerator in Eq.~\eqref{eqn:final_xi}, and we obtain (Appendix \ref{xi-largeNsB})
\begin{equation}\label{eqn:xi-big-Nsb}
 \xi \approx \frac{\displaystyle(1-e^{-2s\beta})}{\displaystyle(N+1)/N-\exp[-NU(1-e^{-2s})]}.
\end{equation}

\subsection{Simulations}

\begin{figure}[htbp]
\centering
\includegraphics[width=120mm]{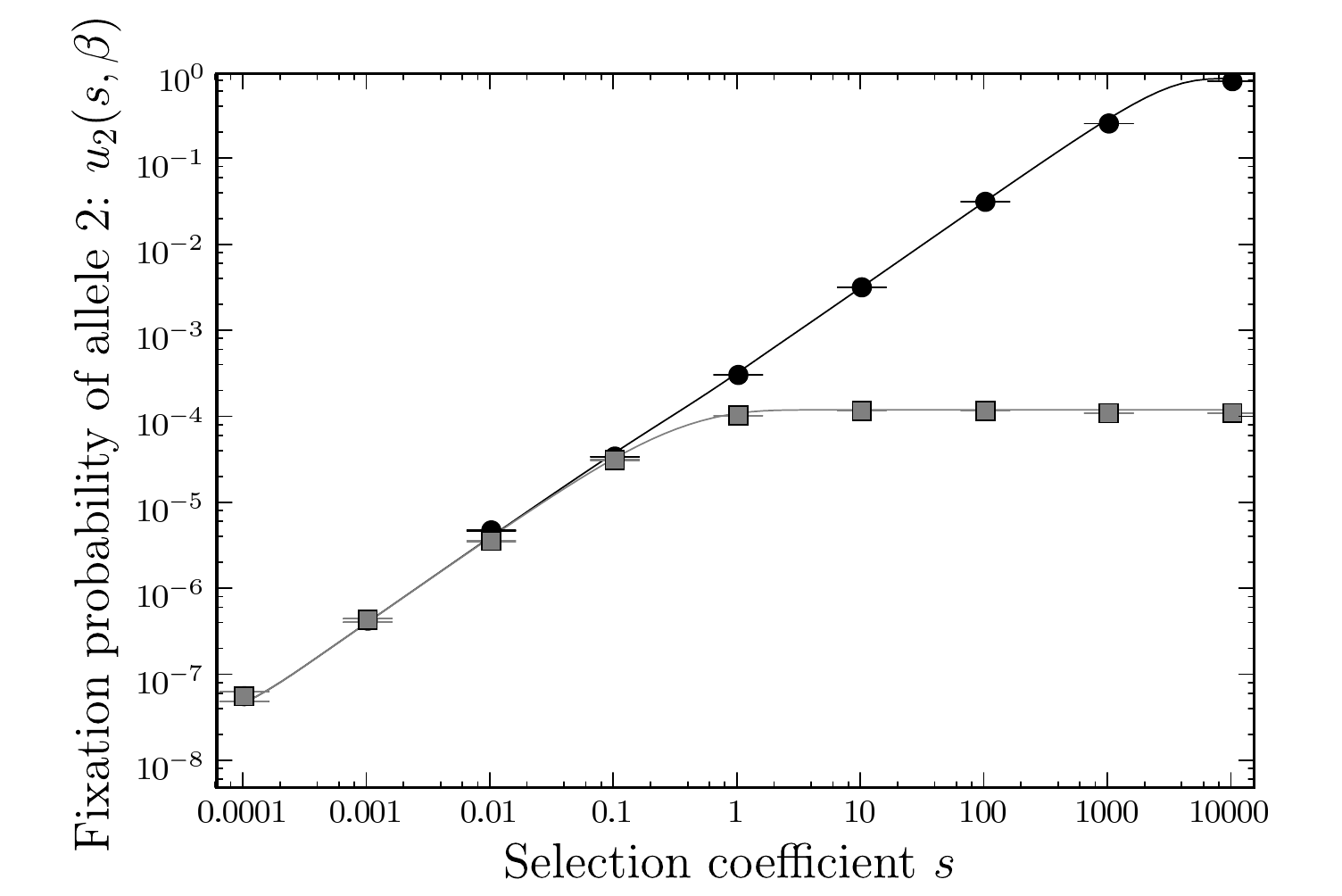}
\caption{Fixation probability of allele 2 ($u_2$) 
vs. the selection coefficient $s$.  Black is for $u_2(s,\beta)$, grey is for $u_2(s,0)$.  Solid lines are predictions according to Eq.~\eqref{u2generalT} and~\eqref{u2fullneutralT}, data points are for simulations with $10^9$ repeats.  $N=10^4$, $U=\frac{8}{19}10^{-8}$, $\beta=0.00019$, $T=5 \times 10^5$.  Error bars are standard errors.}
\label{fig:results}
\end{figure}

\begin{figure}[htbp]
\centering
\includegraphics[width=120mm]{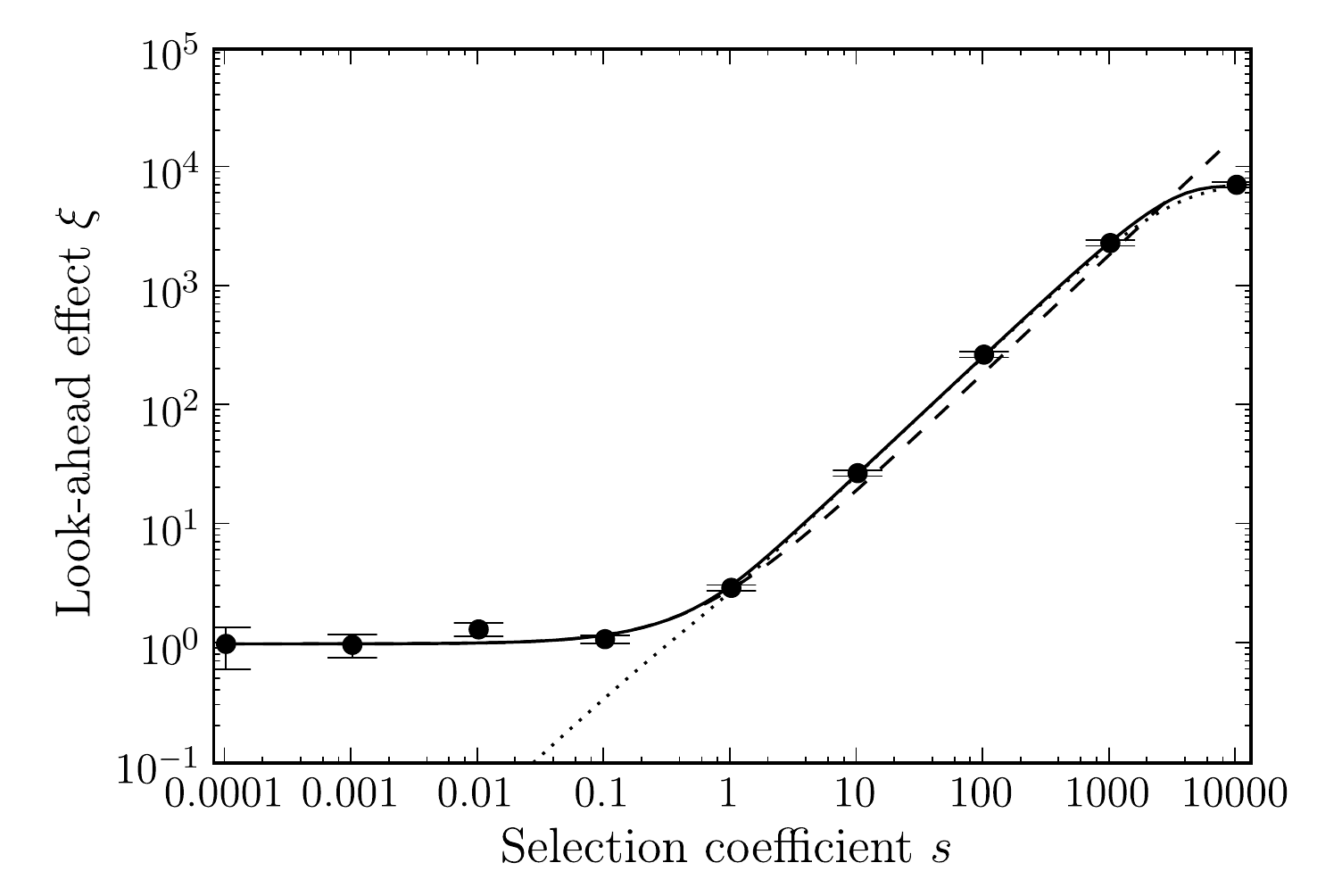}
\caption{Look-ahead effect ($\xi$) due to phenotypic mutations vs. the selection coefficient $s$.  The solid line is for Eq.~\eqref{eqn:final_xi}, dashes are for Eq.~\eqref{eqn:xi-approx1}, dots are for Eq.~\eqref{eqn:xi-big-Nsb}, and  data points are for simulations with $10^9$ repeats.  $N=10^4$, $U=\frac{8}{19}10^{-8}$, $\beta=0.00019$, $T=5 \times 10^5$.  Error bars are standard errors.}
\label{fig:xi}
\end{figure}

We confirmed our analytic results for the fixation probabilities $u_2(s,\beta)$ and $u_2(s,0)$ by numerical simulation, for different values of $s$ (Figure \ref{fig:results}).  With a population size $N=10^4$, the effect of phenotypic mutations can be seen for $s > 0.1$, and  increases for larger $s$.  For $s< 0.1$, the effect is too small and the intermediate allele is effectively neutral, meaning the fixation of allele 2 depends on the random fixation of the neutral allele 1.  The look-ahead effect, $\xi$, shows the simulation results compared to Equations \eqref{eqn:final_xi}, \eqref{eqn:xi-approx1} and \eqref{eqn:xi-big-Nsb}.  Figure \ref{fig:xi} shows the magnitude of the look-ahead effect for the same parameter settings. For large $s$, the look-ahead effect can inflate the probability of fixation of allele 2 by several orders of magnitude.

We also display the different analytic expressions for $\xi$ in Figure \ref{fig:xi}. The approximation \eqref{eqn:final_xi}, derived in the limit $T\rightarrow\infty$, works well for all values of $s$. The approximation \eqref{eqn:xi-approx1}, derived for small $s\beta$, captures correctly the magnitude of $s$ at which the look-ahead effect starts to operate, i.e., $s\gtrsim1/(N\beta)$. Similarly, approximation \eqref{eqn:xi-big-Nsb}, valid for $Ns\beta\gg1$, approximates $\xi$ well for larger $s$.

\begin{figure}[htbp]
\centering
\includegraphics[width=120mm]{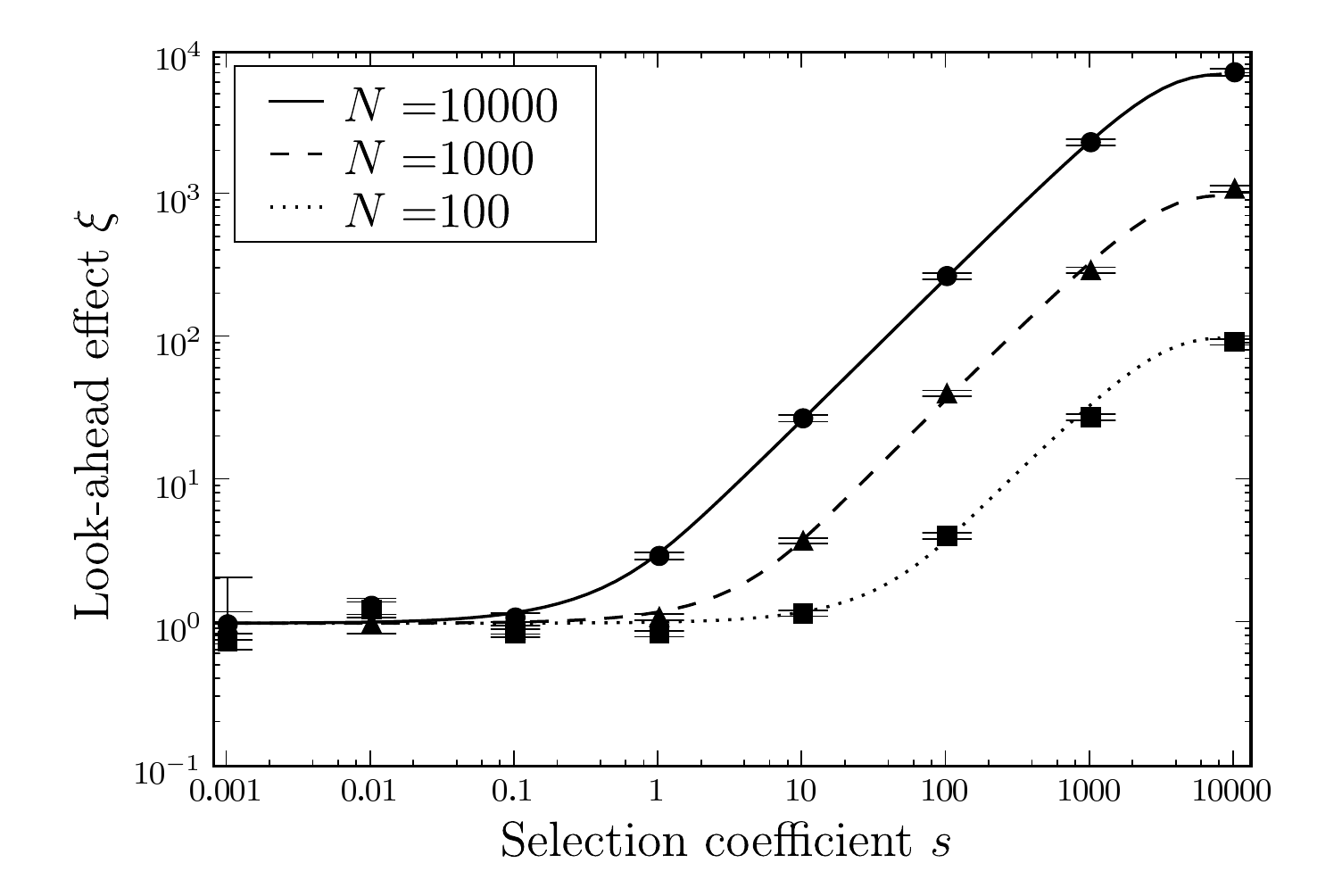}
\caption{Look-ahead effect ($\xi$) due to phenotypic mutations vs. the selection coefficient $s$ for different population sizes ($N$).  Solid lines are from Equation \eqref{eqn:final_xi}, 
data points are for simulations with $10^8$ repeats.  $U=\frac{8}{19}10^{-8}$, $\beta=0.00019$, $T=5 \times 10^5$.  Error bars are standard errors.}
\label{fig:xi-changingN}
\end{figure}

Figure \ref{fig:xi-changingN} shows $\xi$ for different population sizes.  As expected from the condition $s\gtrsim1/(N\beta)$, the look-ahead effect will work with smaller selection coefficients $s$ in larger populations. For large $s$, $\xi$ saturates at approximately $N$.

\begin{figure}[htbp]
\centering
\includegraphics[width=120mm]{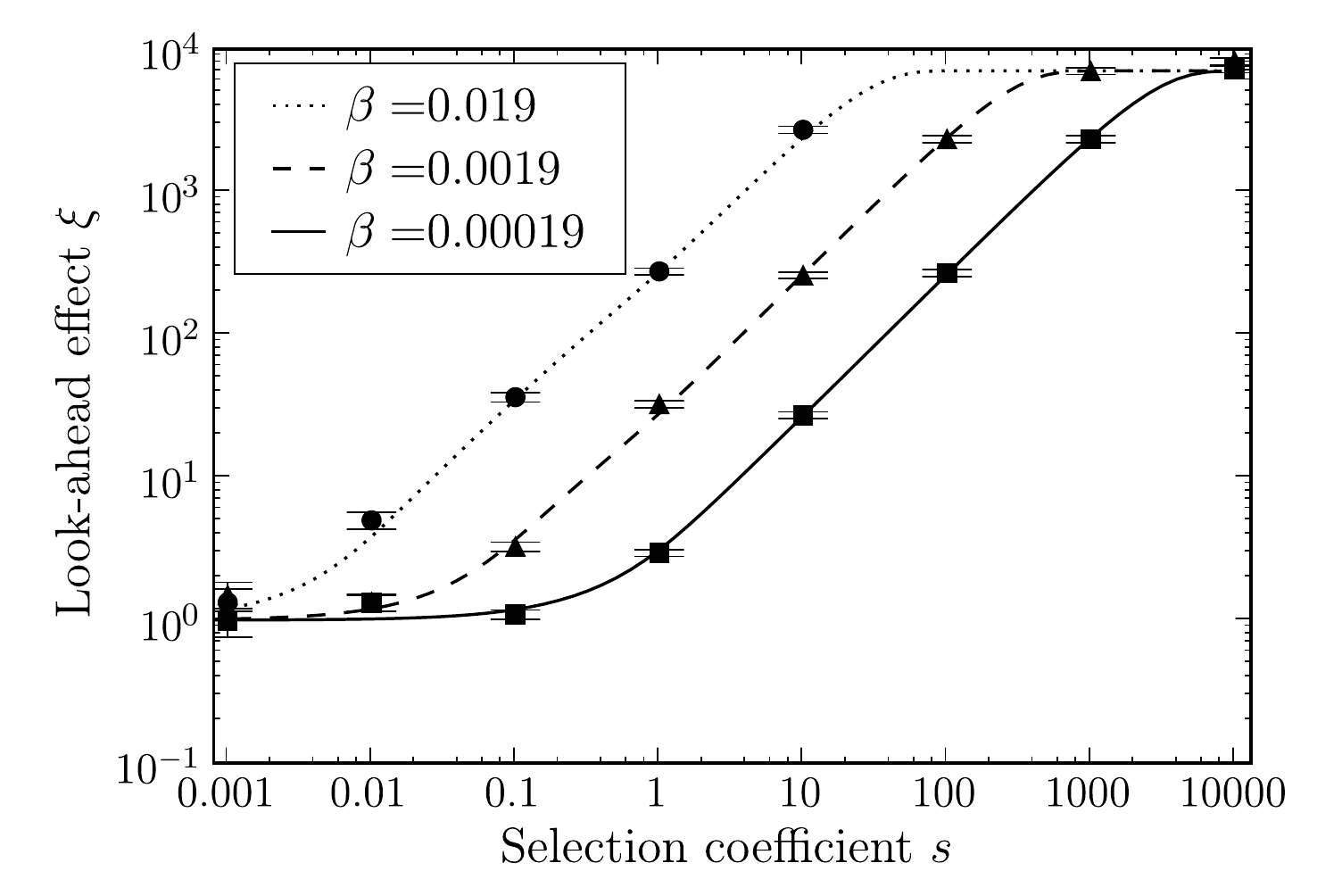}
\caption{Look-ahead effect ($\xi$) due to phenotypic mutations vs. the selection coefficient $s$ for different phenotypic error rates ($\beta$).  Solid lines are from Equation \eqref{eqn:final_xi}, 
data points are for simulations with $10^8$ repeats.  $N=10^4$, $U=\frac{8}{19}10^{-8}$, $T=5 \times 10^5$.  Error bars are standard errors.}
\label{fig:xi-changinglambda}
\end{figure}

We studied the effect of different values of the phenotypic error rate $\beta$ (Fig.~\ref{fig:xi-changinglambda}).  As the error rate $\beta$ increases, the look-ahead effect $\xi$ increases by the same order of magnitude.  For a very high phenotypic error rate of $\beta=0.019$, the look-ahead effect is present for very small values of $s$. However, such a high error rate is likely to be severely detrimental, and in our model we do not take into account the loss of overall fitness for increasing phenotypic error rates.  Conversely for smaller $\beta$,  the look-ahead effect is restricted to large $s$.

\clearpage
\section{Discussion}

We have described a model demonstrating the consequences of positive phenotypic mutations on the evolution of a single gene.  We have compared numerical simulations with the analytical approximations and found them to be in good agreement.  
When phenotypic mutations exert an effect on fitness, selection can operate on the intermediate allele of a complex trait, which otherwise (without phenotypic mutations) would be neutral.  We refer to selection for the intermediate allele as the \emph{look-ahead effect}, because this effect allows evolution to select for sequences not yet in the genome.  

The approximation for small $s\beta$, Eq.\eqref{eqn:xi-approx1}, shows most clearly the relationship between the parameters.  The look-ahead effect is proportional to $N$, $s$, and $\beta$, and sets in when $N$ is on the order of $1/(s\beta)$. For large $Ns\beta$, the look-ahead effect saturates. The asymptotic value of $\xi$ is approximately $N$ for $NU\ll 1$. Therefore,  large populations have two advantages over small populations in terms of the look-ahead effect: the effect sets in for smaller values of $s$, and saturates at a larger asymptotic value $\xi$. Of course, even in the absence of the look-ahead effect, larger populations can more easily traverse multiple local fitness peaks \citep{Weinreich-2005}. 

Because the selection coefficient $s$ depends on the environment, a valid question is how often does $s$ reach sufficiently high levels so that the look-ahead effect can operate.  For microbial species such as bacteria, sufficiently large $s$ should be reasonably common.  
Many bacteria experience highly fluctuating \citep{Smit-2001} and structured \citep{Baquero-1997} environments, where growth is limited by the lack of a key trait.  
An obvious and extreme example is antibiotic resistance.  
Evolving a defense against an antibiotic molecule can involve only a few amino acids \citep{Palzkill-1994}, and those individuals that can generate an enzyme capable of degrading the antibiotic, even if briefly or weakly, will have a very large fitness increase.  
In fact, if the efficacy of the antibiotic is 100\% on susceptible genotypes, a mutation providing only moderate resistance has an infinite selective advantage. 
And even for very small antibiotic concentrations, mutants differing by only two amino acids at a single $\beta$-lactamase gene can be selected effectively \citep{Baquero-1997c,Baquero-1998}. Thus, bacteria may frequently experience  environments in which a large fitness increase (large $s$) is only a few mutations away.  
Similarly, in bacteriophages, selective coefficients $s$ of 10 or more are not
uncommon, even for individual mutations \citep{Bull-2000}.

Our work is entirely theoretical, but we expect that it will be possible to experimentally verify our predictions in future work. For experimentally observing the look-ahead effect, we would need a system where $s$ and $N$ are both large, while $\beta$ (the phenotypic mutation rate) can be modified.  The values of both $N$ and $s$ used in this work are well within biologically realistic ranges achievable in a microbiological laboratory.   Conditions for large $s$ may be created with e.g. antibiotic resistance, which is a common laboratory workhorse. Unfortunately, many antibiotics function by reducing translation fidelity \citep{Ogle-2005}, and thus would conflate $s$ and $\beta$. Changing $\beta$ could involve a mutated ribosome.  Ribosomes appear to be optimized for accurate and efficient translation of mRNA \citep{Baxter-Roshek-2007}, and several examples of altered ribosome fidelity exist, both increasing \citep{Vila-Sanjurjo-2003} and decreasing fidelity \citep{Friedman-1968}. Specific regions of the ribosome rRNA sequence have been identified as influencing fidelity \citep{O'Connor-1997}, and various agents can reduce fidelity, e.g., streptomycin, magnesium, and ethanol \citep{Friedman-1968}.  Few mutations may be sufficient to alter the fidelity of a ribosome, for example, a single mutation in the S5 ribosomal protein in \emph{E. coli} increases frameshifting and nonsense mutations \citep{Kirthi-2006}. In yeast, mutations in the 18S RNA have been found that both increase and decrease translational fidelity \citep{Konstantinidis-2006}.   

In this work, we have calculated the look-ahead effect from a comparison between the two cases of $\beta > 0$ and $\beta=0$. The latter may not be experimentally possible; any experiment will likely compare two different positive values of $\beta$. Nevertheless, Figure \ref{fig:xi-changinglambda} shows that a larger look-ahead effect can be achieved with a higher $\beta$,  where increasing $\beta$ by one order of magnitude both increases the look-ahead effect by an order of magnitude and lowers the smallest $s$ where an effect is observed.  Of course, our model does not take into account the loss of fitness or other confounding effects from a higher phenotypic mutation rate.  Thus, a balance must be found in having two different values of $\beta$ that are different enough to measure, while at the same time minimizing the confounding effects.  The most obvious consequence of increasing the phenotypic mutation rate is that overall fitness may be reduced, for example in \emph{E. coli}, where a higher translational error rate activates stress responses \citep{Fredriksson-2007}, or in mouse, where such errors are implicated in  neurodegeneration caused by misfolded proteins that aggregate \citep{Lee-2006}.  Increasing translational fidelity may not come without fitness cost either. The hyperaccurate mutations  in the 18S RNA in yeast \citep{Konstantinidis-2006} cause an increase in oxidative stress. This observation  suggests that cells consume more energy to achieve hyperaccuracy.  It may also partially explain why the phenotypic error rate is much higher than the genotypic error rate, as there is possibly a direct disadvantage in reducing the phenotypic error rate, rather than only reducing the selective advantage that occurs if the phenotypic error rate is reduced, as discussed in \citet{Buerger-2006}.  

\citet{Buerger-2006} asked whether evolution has selected for the current phenotypic error rate, which does not differ significantly between eukaryotes and prokaryotes \citep{Ellis-1982,Shaw-2002}  even though the source of errors is different.  They suggested that the increase in fitness becomes incrementally smaller for improvements to transcription and translation fidelity.  We would like to speculate that the phenotypic error rate is on the border between minimal costs (of e.g.\ misfolded proteins) and maximum payoff (via the look-ahead effect).

The goal of our analysis was to demonstrate that the look-ahead effect is theoretically possible, and as such, we intentionally excluded confounding factors for the sake of clarity.  There are several aspects not considered in our model that may play important roles.  For example, in this work we did not consider the expression level.  For low expressed genes, the mutation from allele 1 to allele 2 will occur less frequently compared to highly expressed genes.  However, if allele 2 is produced it will be at a higher concentration (of allele 2 mutant proteins in a population of allele 1 proteins), as the overall copy number of allele 1 is low.  
This difference in expression levels is likely reduced in a large population, where beneficial mutations occur with sufficient frequency.  Another factor related to the expression level is translational robustness.  It has been proposed that highly expressed genes are under selection to properly fold despite phenotypic mutations, and consequently evolve slower \citep{Drummond-2005a, Wilke-2006}.  If a gene is robust to translational errors, then it can tolerate a larger variety of mutations, of which some may be intermediates to a new adaptive multi-residue trait.  Thus, translational robustness may increase the  sequences available for experimentation at the phenotypic level. However, if the intermediate allele is itself not robust to errors in translation, then it will not be neutral, and may be selected against.  
The location of the protein trait will also influence the viability of the intermediate allele: mutations near the surface of the protein are less likely to disrupt the protein compared to mutations in the core \citep{Tokuriki-2007}.

In conclusion, we propose that organisms can experiment with protein sequences that are mutationally close to the current sequence, but not yet in the genome.  This effect allows selection for intermediates of complex traits, opening up a more direct route to the trait and thus reducing the time needed for fixation in the population.

\section{Acknowledgements}

D.J.W. would like to thank January Weiner for stimulating discussions and Maya Amago for helpful suggestions.  D.J.W. and E.B.B. were supported by an HFSP program grant.  C.O.W. was supported by NIH grant AI 065960.

\appendix
\section{Appendix}\label{Appendix}

\renewcommand{\theequation}{A\arabic{equation}}
\setcounter{equation}{0}

Here, we present the details of our analytic derivations.

\subsection{Probability of fixation}
\label{probability-of-fixation}

According to \citep{Kimura-1962}, the probability of fixation $u(s)$ of a single allele with selection coefficient $s$ is given by
\begin{equation}
  u(s) = \frac{1-e^{-2s}}{1-e^{-2Ns}}.
\end{equation}
For $s\lesssim 1/N$, this expression simplifies to
\begin{equation}\label{ufixsmall}
  u(s) = \frac{1}{N} + \frac{N-1}{N} s + {\cal O}(s^2),
\end{equation}
whereas for $Ns\gg1$, this expression simplifies to
\begin{equation}\label{ufixlarge}
  u(s) \approx 1-e^{-2s}.
\end{equation}

\subsection{A single allele drifting to fixation or loss}\label{single-allele-loss}

We first consider a single allele with selective advantage $s$ drifting to fixation or extinction, and ask how many mutations this allele generates until it is either fixed or lost. We will treat these two cases separately. Let $n_\text{fix}(s)$ be the expected number of mutations generated while the allele drifts to fixation, and let $n_\text{loss}(s)$ be the expected number of mutations generated while the allele drifts to extinction. We calculate these two quantities using diffusion theory, by integrating the sojourn times of the allele over all frequencies.

For an allele with selective coefficient $s$ and starting at frequency $p=1/N$, \citet{Nagylaki-1974} calculated its mean sojourn time $\tau(y)$ between frequencies $y$ and $y+dy$ as
\begin{equation}\label{sojourn-basic}
  \tau(y) = 2[V(y)G(y)]^{-1}[u_\text{loss}(1/N)g(0,y)\theta(1/N-y)+u_\text{fix}(1/N)g(y,1)\theta(y-1/N)].
\end{equation}
Here,
\begin{align}
  V(y)G(y)&=y(1-y)e^{-2Nsy}/N,\\
  g(a,b) &= \frac{e^{-2Nsa}-e^{-2Nsb}}{2Ns},\\
  u_\text{loss}(p) &= \frac{e^{-2Nsp}-e^{-2Ns}}{1-e^{-2Ns}},\\
  u_\text{fix}(p) &=1-u_\text{loss}(p)=\frac{1-e^{-2Nsp}}{1-e^{-2Ns}},
\end{align}
and $\theta(z)$ is the Heaviside step function. We want to integrate expressions involving $\tau(y)$ from $y=0$ to $y=1$. Since $y=1/N$ corresponds to a single copy of the allele that drifts to fixation, values of $y$ less than $1/N$ are not relevant for our analysis. Therefore, we discard the term proportional to $\theta(1/N-y)$ in Eq.~\eqref{sojourn-basic}, and use in what follows
\begin{equation}\label{sojourn-simplified}
  \tau(y) = 2u_\text{fix}(1/N)g(y,1)/[V(y)G(y)]  \qquad \text{for $y>1/N$}.
\end{equation}

\subsection{Number of mutations conditional on fixation}
\label{number-of-mutations-conditional-on-fixation}

For the sojourn time conditional on fixation, $\tau_\text{fix}(y)$, \citet{Nagylaki-1974} finds
\begin{equation}
  \tau_\text{fix}(y) = \tau(y)u_\text{fix}(y)/u_\text{fix}(p).
\end{equation}
Using this expression, we have
\begin{equation}\label{nfix-def}
  n_\text{fix}(s) = NU\int_{1/N}^1 \tau_\text{fix}(y) ydy.
\end{equation}
Plugging the expressions for $V(y)G(y)$, $g(a,b)$, $u_\text{fix}(p)$, and $\tau(y)$ into $\tau_\text{fix}(y)$, we arrive at
\begin{equation}
  \tau_\text{fix}(y) = \frac{1}{s(1-e^{-2Ns})} \frac{(1-e^{-2Nsy})(1-e^{-2Ns(1-y)})}{y(1-y)}.
\end{equation}
This expression corresponds to the one by \citet{Ewens-1973}. Note that $y\tau_\text{fix}(y)\rightarrow0$ for $y\rightarrow 0$. Therefore, we can extend the lower limit of integration to 0 in Eq.~\eqref{nfix-def}, and rewrite $n_\text{fix}(s)$ as
\begin{equation}\label{nfix}
  n_\text{fix}(s) = \frac{NU}{s(1-e^{-2Ns})} I(2Ns)
\end{equation}
with
\begin{equation}
  I(a) = \int_0^1 \frac{(1-e^{-a y})(1-e^{-a(1-y)})}{1-y}dy.
\end{equation}
The integral $I(a)$ can be rewritten as 
\begin{equation}\label{Iofa}
  I(a) = \gamma - \Ei(-a)+\ln(a)+e^{-a}[\gamma-\Ei(a)+\ln(a)],
\end{equation}
where $\gamma\approx0.5772$ is the Euler-Mascheroni constant and $\Ei(z)$ is the exponential integral,
\begin{equation}
  \Ei(z) = -\int_{-z}^\infty \frac{e^{-t}}{t} dt.
\end{equation}
For $s\lesssim 1/N$, we find
\begin{equation}\label{nfixsmall}
  n_\text{fix}(s) = N^2 U + {\cal O}(s^2).
\end{equation}
For $Ns\gg 1$, we obtain the asymptotic expansion
\begin{equation}\label{nfixlarge}
  n_\text{fix}(s) \approx \frac{N U}{s} [\ln(2Ns)+\gamma],
\end{equation}
using \citet{Abramowitz-1964} 5.1.51,
\begin{equation}\label{Eiasymp}
  \Ei(-z)\sim -\frac{e^{-z}}{z}\Big(1-\frac{1}{z}+\frac{2}{z^2}-\frac{6}{z^3}\Big) \qquad \text{for large $z$}.
\end{equation}

\subsection{Number of mutations conditional on extinction}
\label{number-of-mutations-conditional-on-extinction}
For the sojourn time conditional on extinction, $\tau_\text{loss}(y)$, \citet{Nagylaki-1974} finds
\begin{equation}
  \tau_\text{loss}(y) = \tau(y)u_\text{loss}(y)/u_\text{loss}(p).
\end{equation}
Using this expression, we have
\begin{equation}
  n_\text{loss}(s) = NU\int_{1/N}^1  \tau_\text{loss}(y) ydy.
\end{equation}
Plugging the expressions for $V(y)G(y)$, $g(a,b)$, $u_\text{loss}(p)$, and $\tau(y)$ into $\tau_\text{loss}(y)$, we find
\begin{equation}
  \tau_\text{loss}(y) = \frac{1}{s(1-e^{-2Ns})}  \frac{e^{2s}-1}{1-e^{-2(N-1)s}}\frac{(e^{-2Nsy}-e^{-2Ns})(1-e^{-2Ns(1-y)})}{y(1-y)}.
\end{equation}
We rewrite $n_\text{loss}$ as
\begin{equation}
  n_\text{loss} =  \frac{NU}{s(1-e^{-2Ns})}  \frac{e^{2s}-1}{1-e^{-2(N-1)s}} J(N,s)
\end{equation}
with
\begin{equation}
  J(N,s) = \int_{1/N}^1\frac{(e^{-2Nsy}-e^{-2Ns})(1-e^{-2Ns(1-y)})}{1-y} dy.
\end{equation}
The integral can be rewritten as 
\begin{equation}
  J(N,s) = -2e^{-2Ns}\big(\gamma-\Chi[2(N-1)s]+\ln[2(N-1)s]\big),
\end{equation}
where $\Chi(z)$ is the hyperbolic cosine integral,
\begin{equation}
  \Chi(z) = \gamma+\ln(z)+\int_0^z \frac{\cosh(t)-1}{t} dt.
\end{equation}
For $s\lesssim 1/N$, we find
\begin{equation}\label{nlosssmall}
  n_\text{loss}(s) = (N-1) U + {\cal O}(s^2).
\end{equation}
For $Ns\gg 1$, we obtain the asymptotic expansion
\begin{equation}\label{nlosslarge}
  n_\text{loss}(s) \approx \frac{U}{2s^2} (1-e^{-2s}),
\end{equation}
using
\begin{equation}
  \Chi(z)\approx \frac{\Ei(z)}{2} \approx \frac{e^z}{2z}\qquad\text{for large $z$.}
\end{equation}
[This expansion follows directly from the definitions of $\Chi(z)$, $\cosh(z)$, and $\Ei(z)$.]

\subsection{Number of mutations within a given time interval}\label{mutations-given-time-interval}

We now extend the derivations in Section \ref{number-of-mutations-conditional-on-fixation} to calculate the number of mutations to allele 2 generated within a certain time interval $T$, conditional on fixation of allele 1. We assume that $T$ is sufficiently large so that allele 1 has time to reach fixation within this interval. We only consider the case conditional on fixation because no new mutations are generated once allele 1 has gone extinct.

We calculate $n(s)=n_\text{fix}(s)+n_\text{T}(s)$, where $n_\text{T}(s)$ is the total number of mutations generated once the first mutation has reached fixation. We have
\begin{equation}\label{nT}
  n_\text{T}(s) = NU[T-t_\text{fix}(s)],
\end{equation}
where $t_\text{fix}(s)$ is the time to fixation of a mutation with selective advantage $s$. This time is given by the integral over all sojourn times,
\begin{equation}
  t_\text{fix}(s) = \int_0^1 \tau_\text{fix}(y) dy = \frac{I_2(2Ns)}{s(1-e^{-2Ns})}
\end{equation}
with
\begin{equation}
  I_2(a) = \int_0^1 \frac{(1-e^{-a y})(1-e^{-a(1-y)})}{y(1-y)}dy.
\end{equation}
A partial fraction decomposition of the integrand reveals that $I_2(a)=2I(a)$, and thus we have
\begin{equation}
  t_\text{fix}(s) =  \frac{2I(2Ns)}{s(1-e^{-2Ns})}
\end{equation}
Combining this result with Eqs.~\eqref{nfix} and~\eqref{nT}, we find
\begin{align}
  n(s)&=n_\text{fix}(s)+n_\text{T}(s) = NU\Big[T- \frac{I(2Ns)}{s(1-e^{-2Ns})}\Big]\notag\\
     &= NUT- n_\text{fix}(s).
\end{align}
Note that $n(s)=n_\text{fix}(s)$ for $T=t_\text{fix}(s)$.

For $s\lesssim 1/N$, we find
\begin{equation}\label{nsmall}
  n(s) = N U(T-N) + {\cal O}(s^2).
\end{equation}
For $Ns\gg 1$, using Eqs.~\eqref{Iofa} and~\eqref{Eiasymp}, we obtain the asymptotic expansion
\begin{equation}\label{nlarge}
  n(s) \approx N U\Big(T- \frac{\ln(2Ns)+\gamma}{s}\Big).
\end{equation}

\subsection{$\xi$ for $s\beta\ll1$}\label{xi-smalls}

From Eq.~\eqref{eqn:xi}, using Eqs.~\eqref{nlosssmall}, \eqref{nsmall}, and \eqref{ufixsmall}, we obtain to first order in $s\beta$
\begin{equation}
  \xi \approx 1 +\frac{e^{-NUu(s)}-e^{-NU(T-N)u(s)}}{u_2(s,0)} s\beta + {\cal O}(s^2\beta^2).
\end{equation}
If further $NU(T-N)u(s)\ll 1$, we obtain
\begin{equation}
  \xi \approx 1+N(1-N/T) s\beta + {\cal O}(s^2\beta^2), 
\end{equation}
and for $T\rightarrow\infty$, 
\begin{equation}
  \xi \approx 1+Ns\beta + {\cal O}(s^2\beta^2).
\end{equation}

\subsection{$\xi$ for $N s \beta\gg 1 $}\label{xi-largeNsB}
For $Ns\beta\gg1$, only the first term contributes to Eq.~\eqref{u2generalT}, and we obtain from 
Eqs.~\eqref{nlarge} and \eqref{ufixlarge}

\begin{equation}
  u_2(s,\beta) = (1-e^{-2s\beta})\Big[1-\exp\Big(-NU\Big[T-\frac{\ln(2Ns\beta)+\gamma}{s\beta}\Big]\big(1-e^{-2s}\big)\Big)\Big].
\end{equation}
Likewise, in this limit we can simplify Eq.~\eqref{u2fullneutralT} to
\begin{equation}
  u_2(s,0) = \frac{N+1}{N}-\exp[-NU(T-N)(1-e^{-2s})]/N-\exp[-NU(1-e^{-2s})], 
\end{equation}
giving 
\begin{equation}
 \xi \approx \frac{\displaystyle(1-e^{-2s\beta})\Big[1-\exp\Big(-NU\Big[T-\frac{\ln(2Ns\beta)+\gamma}{s\beta}\Big]\big(1-e^{-2s}\big)\Big)\Big]}{\displaystyle(N+1)/N-\exp[-NU(T-N)(1-e^{-2s})]/N-\exp[-NU(1-e^{-2s})]}.
\end{equation}
Furthermore, for $T\rightarrow\infty$, this expression simplifies to 
\begin{equation}
 \xi \approx \frac{\displaystyle(1-e^{-2s\beta})}{\displaystyle(N+1)/N-\exp[-NU(1-e^{-2s})]}.
\end{equation}
If $NU\ll 1$, then $\xi\rightarrow N$ in the limit $s\rightarrow\infty$.

\bibliographystyle{genetics}
\bibliography{references}

\end{document}